\documentclass[prd,showpacs,nofootinbib,onecolumn,amsmath,amsfonts,amssymb]{revtex4} %,nofootinbib

\usepackage{amsmath}
\usepackage{amssymb}
\usepackage{epsfig}

\usepackage{graphicx}
\usepackage{stmaryrd}
\usepackage{bm}
\usepackage{color}

\newcommand{\AddrMPP}{%
Max-Planck-Institut f\"ur Physik (Werner Heisenberg
Institut), F\"ohringer Ring 6, 80805 M\"unchen, Germany}

\begin{document}

\title{Constraints on very light sterile neutrinos from $\theta_{13}$-sensitive reactor experiments}

\author{Antonio Palazzo}
\affiliation{\AddrMPP}

\date{\today}

\begin{abstract}

Three dedicated reactor experiments, Double Chooz, RENO and Daya Bay, have recently performed a 
precision measurement of the third standard mixing angle $\theta_{13}$ exploiting
a multiple baseline comparison of  $\nu_e \to \nu_e$ disappearance driven by the atmospheric
mass-squared splitting. In this paper we show how the same
technique can be used to put stringent limits on the oscillations of the electron neutrino 
into a fourth very light sterile species (VLS$\nu$) characterized by a
mass-squared difference lying in the range $[10^{-3} - 10^{-1}]\,\mathrm{eV}^2$. 
We present accurate constraints on the admixture $|U_{e4}|^2$  obtained by a 4-flavor analysis
of the publicly available reactor data. In addition, we show that the estimate of $\theta_{13}$
obtained by the combination of the three reactor experiments is rather robust and substantially independent of the 4-flavor-induced perturbations provided that  the new
mass-squared splitting  is not too low ($\gtrsim 6 \times10^{-3}\,\mathrm{eV}^2$). 
We briefly comment on the possible impact of VLS$\nu$'s on the rest of the neutrino oscillation
phenomenology and emphasize their potential role in the cosmological ``dark radiation'' anomaly.

\end{abstract}


\maketitle

\section{Introduction}

While the 3-flavor framework has been solidly established as a standard paradigm,
it may not constitute the ultimate description of the neutrino oscillation phenomena. 
Striking deviations from the standard picture arise in the presence of new additional light neutrinos 
if these are mixed with the ordinary species. From the LEP measurement of the invisible decay
width of the Z boson~\cite{ALEPH:2005ab}, we know that there are only three light neutrinos which participate to the
electroweak interactions. Therefore, the new putative
light neutral fermions must be ``sterile'', i.e. singlets of the Standard Model gauge group,
to be contrasted with the ordinary ``active'' neutrino species, which are members of weak isospin doublets.

The theory is not able to provide solid information on the sterile neutrino mass-mixing properties, 
which ultimately need to be constrained (and hopefully determined) by the experiments. Accordingly, any new opportunity
to enlarge our knowledge of the sterile mass-mixing parameters should not remain unexploited. With this attitude, 
in the present paper we will investigate very light sterile neutrinos (VLS$\nu$'s), i.e. singlet neutral fermions separated from the active species by a mass-squared splitting in the range $|\Delta m^2_{14}| \equiv | m^2_4 - m^2_1| \in [10^{-3} -10^{-1}]~\mathrm{eV}^2$. In particular, we will focus on the constraints attainable on the admixture of the electron neutrino with the new mass eigenstate $\nu_4$, which is parametrized by the lepton matrix element $|U_{e4}|^2$.

Additional phenomenological motivation for investigating VLS$\nu$'s comes from cosmology. In fact, for mass-squared splittings
in the range $[10^{-3} - 10^{-1}]$~eV$^2$ and sufficiently small admixtures  $|U_{e4}|^2 \lesssim {\mathrm {few}}
 \times 10^{-2}$, a fourth sterile neutrino is only partially thermalized in the early universe and thus provides
a fractional contribution $\Delta N_{\mathrm {eff}} \in [0, 1]$ to the number of effective 
relativistic degrees of freedom (see~\cite{Hannestad:2012ky,Mirizzi:2013kva}), in agreement with the latest
cosmological measurements~\cite{Ade:2013zuv}, which indicate $\Delta N_{\mathrm {eff}} = 0.62^{ +0.50}_{-0.48}$
(95\% C.L.). Moreover, the absolute neutrino mass content implied by a 3+1 scheme involving a VLS$\nu$  is 
very small and fully compatible with the existing upper limits (see for example~\cite{Giusarma:2013pmn}).
These features render VLS$\nu$'s  excellent candidates to explain the ``dark radiation" anomaly as well as the recent hints of a hot dark matter component requiring a particle mass in the sub-eV range%
 %%%%%%%%%%%%%%%%%%%%%%%%%%%%%%%%%%%%%%%%%%%%%%%%%%%%%%%%%%%%%
\footnote{The same properties are not shared by the heavier sterile neutrinos with mass of $\sim 1~$eV invoked to interpret the anomalies recorded at very short baseline accelerator, reactor and gallium experiments (see~\cite{Kopp:2013vaa,Giunti:2013aea} for updated analyses). Indeed,  in  this case, one expects full thermalization~\cite{Hannestad:2012ky,Mirizzi:2013kva} ($\Delta N_{\mathrm {eff}} = 1$) and a larger absolute neutrino mass content, both in tension with current cosmological constraints~\cite{Hamann:2013iba}.}
%%%%%%%%%%%%%%%%%%%%%%%%%%%%%%%%%%%%%%%%%%%%%%%%%%%%%%%%%%%%%  
\cite{Wyman:2013lza,Hamann:2013iba,Burenin:2013wg}.

The existing constraints on the electron neutrino admixture with VLS$\nu$'s are limited to the region of  $|\Delta m^2_{14}| \gtrsim  {few} \times 10^{-2}\,  \mathrm{eV}^2$ and were obtained by reactor experiments with baselines of a few tens of meters: Bugey-3~\cite{Declais:1994su}, G\"osgen~\cite{Zacek:1986cu} and Krasnoyarsk~\cite{Vidyakin:1987ue,Vidyakin:1994ut}. It is thus natural to expect that lower values of $\Delta m^2_{14}$ may be probed through reactor experiments with longer baselines 
and in particular by the novel $\theta_{13}$-dedicated experiments Double Chooz,  RENO and Daya Bay. 
We remind that the measurement of $\theta_{13}$ performed in such setups relies upon the observation 
of a disappearance phenomenon of the electron (anti-)neutrinos driven by oscillations intervening at the atmospheric 
mass-squared splitting
 ($|\Delta m^2_{13}| \sim 2.4 \times 10^{ -3}~$eV$^2$), to which such detectors have been calibrated.
In particular, the far site baselines (ranging from 1 to 2$~$km) have been chosen in order to maximize the oscillation 
effects induced by such a frequency, while the near sites have been fixed at a few hundred meters
from the reactors, where the (3-flavor) oscillation effects are very small. 
 In the presence of an additional oscillation frequency, as in the case of sterile neutrinos, it is 
expected a sensitivity of the same setups to mass-squared splittings in the range $|\Delta m^2_{14}| \in [10^{-3} - 10^{-1}]~\mathrm{eV}^2$, 
as these induce non negligible effects both at the far and at the near sites. Indeed, this potentiality has been already recognized in previous
works~\cite{Mikaelyan:1998yg,Kopeikin:2003uu,deGouvea:2008qk,Bora:2012pi,Bergevin:2013nea}, in which sensitivity studies have been performed. As real data from the $\theta_{13}$-dedicated reactor experiments have become now available, 
it is timely to exploit  them to constrain a window of the parameter space so far unexplored.

The full potential of $\theta_{13}$-sensitive reactor experiments in constraining VLS$\nu$'s properties arises from the exploitation of both
their total rate and spectral energy information.  Unfortunately,  at the moment, the spectral analysis of the data cannot be
made from outside the collaborations. Indeed, while the Double Chooz~\cite{Abe:2011fz,Abe:2012tg} and Daya Bay~\cite{Daya_Bay_NuFact_2013} collaborations have performed such shape analyses, at the moment it is problematic to reproduce
 them accurately due to missing details on various aspects of the experiments. On the other hand, the RENO
experiment has not (yet) performed such kind of analysis.%
%%%%%%%%%%%%%%%%%%%%%%%%%%%%%%%%%%%%%%%%%%%%%%%%%%%
\footnote{A spectral analysis of RENO and Daya Bay has been recently performed in~\cite{Kang:2013gpa}.
However, the authors do not report any information on the details of their analysis.}
%%%%%%%%%%%%%%%%%%%%%%%%%%%%%%%%%%%%%%%%%%%%%%%%%%%
For these reasons, in this work will limit our study to the total rate information. 
We plan to perform a spectral shape analysis in a future work, when more detailed information 
will be released by all the three collaborations.

A further limitation to the usage of these data comes from the uncertainty on the absolute reactor neutrino fluxes. Indeed,
their recent reevaluation performed in~\cite{Mueller:2011nm,Huber:2011wv} has provided values $\sim3.5\%$ larger than the previous accepted estimates~\cite{Vogel:1980bk,VonFeilitzsch:1982jw,Schreckenbach:1985ep,Hahn:1989zr}, resulting in the so called ``reactor anomaly''~\cite{Mention:2011rk} (see also~\cite{Zhang:2013ela}), i.e. a deficit in the total rates observed at all the reactor experiments with detectors placed at short ($< 100$\, m) distances.
At the moment, the origin of such a  discrepancy is  unknown and is matter of intense investigation. The anomaly may involve new physics intervening at short distances but also be related to systematics in the reactor flux calculations. In such a situation, when using reactor neutrino data, it is opportune not to rely upon that part of information extracted with assumptions on the emitted neutrino fluxes.

In spite of all such limitations, the three $\theta_{13}$-dedicated reactor experiments can still provide precious
information on VLS$\nu$'s.
As we will show, interesting constraints can be obtained {\em independently} on any assumption made on the absolute normalization
of the reactor fluxes, by exploiting exclusively the {\em comparison} of the total rates measured at different baselines.
This ``modus operandi'' --- which is exactly the same one adopted for the determination of $\theta_{13}$ ---
will prove very useful, as it will render the results completely independent of the debated 
theoretical determinations of the reactor neutrino fluxes.%
%%%%%%%%%%%%%%%%%%%%%%%%%%%%%%%%%%%%%%%%%%%%%%%%%%%
\footnote{Very recently the multiple-detector technique with baselines of a few meters
has been proposed in~\cite{Heeger:2013ema} for the investigation of the region of $\Delta m^2_{14} \sim 1\, \mathrm{eV}^2$.}
%%%%%%%%%%%%%%%%%%%%%%%%%%%%%%%%%%%%%%%%%%%%%%%%%%%
In addition, this technique will render the results 
independent of any hypothetical sterile neutrino oscillation phenomenon occurring at baselines of
a few meters, as invoked to explain the reactor~\cite{Mention:2011rk} (see also~\cite{Zhang:2013ela})
and gallium~\cite{Abdurashitov:2005tb,Giunti:2010zu} anomalies  (for which higher values of the
mass-squared splitting $|\Delta m^2_{14}| \sim 1~ \mathrm {eV} ^2$ are relevant). 

The  paper is organized as follows. In Sec.~II we outline the basic formulae needed to
describe the vacuum oscillations in the presence of a VLS$\nu$. 
In Sec.~III we present the physics at the basis of reactor neutrinos production and detection processes.
In Sec.~IV we describe the details of the statistical 
analysis of each of the three reactor experiments and of their combination.
In Sec.~V we perform a numerical study within the standard 3-flavor framework, 
which will be preparatory to the full 4-flavor analysis presented in the following section VI.
Finally, in Sec.~VII we trace our conclusions accompanied by a short discussion.

\section{Neutrino oscillations in the presence of a very light sterile species}

In the presence of a fourth light sterile neutrino species, the electron neutrino $\nu_e$ can be expressed as a linear combination of the mass eigenstates $\nu_i$ ($i=1,2,3,4$),  $\nu_e=\sum_iU_{ei}\nu_i$. 
The electron neutrino (or antineutrino) survival probability in vacuum is simply given by the expression
%%%%%%%%%%%%%%%%%%%%%%%%%%%%%%%%%%%%%%%%%%%%%%%%%%%%%
\begin{eqnarray}
P_{ee}
&=& 1 - \sum_{i<j}4|U_{ei}|^2|U_{ej}|^2\sin^2\left(\frac{\Delta m^2_{ij}L}{4E_{\nu}}\right)~,
\label{eq:general}
\end{eqnarray}
%%%%%%%%%%%%%%%%%%%%%%%%%%%%%%%%%%%%%%%%%%%%%%%%%%%%%
where $L$ is the antineutrino propagation distance (baseline), $E_{\nu}$ is the neutrino energy and
$\Delta m^2_{ij}\equiv m_j^2-m_i^2$.
%%%%%%%%%%%%%%%%%%%%%%%%%%%%%%%%%%%%%%%%%%%%%%%%%%%%%
As shown in~\cite{deGouvea:2008qk}, for $L$ and $E_{\nu}$ values of interest, the survival probability can be approximated as
%%%%%%%%%%%%%%%%%%%%%%%%%%%%%%%%%%%%%%%%%%%%%%%%%%%%%
\begin{equation}
P_{ee}\simeq1-4(1-|U_{e3}|^2-|U_{e4}|^2)|U_{e3}|^2\sin^2\left(\frac{\Delta m^2_{13}L}{4E_{\nu}}\right)-4(1-|U_{e4}|^2)|U_{e4}|^2\sin^2\left(\frac{\Delta m^2_{14}L}{4E_{\nu}}\right)~,
\label{eq:degouvea}
\end{equation}
%%%%%%%%%%%%%%%%%%%%%%%%%%%%%%%%%%%%%%%%%%%%%%%%%%%%%
where small terms suppressed by the fourth power of the two matrix elements%
%%%%%%%%%%%%%%%%%%%%%%%%%%%%%%%%%%%%%%%%%%%%%%%%%%%%%
\footnote{As shown in~\cite{deGouvea:2008qk} this term is the only one potentially sensitive to the sign of $\Delta m_{14}^2$.
Our analysis confirms that such a sensitivity is negligible.}  
%%%%%%%%%%%%%%%%%%%%%%%%%%%%%%%%%%%%%%%%%%%%%%%%%%%%%
 ($|U_{e3}|^2|U_{e4}|^2$) or by the solar neutrino splitting have been neglected (see~\cite{deGouvea:2008qk} for a detailed proof).
In this limit, $P_{ee}$ is a function of the absolute value of the two mass-squared differences ($|\Delta m^2_{13}|$ and $|\Delta m^2_{14}|$) and of the moduli of the two elements  ($|U_{e3}|$ and $|U_{e4}|$) of the lepton mixing matrix.

It is convenient to parameterize the mixing matrix U as
%..........................................................................
\begin{equation}
\label{eq:U}
U =  A R_{14} R_{13} R_{12}, 
\end{equation} 
%..........................................................................
where $R_{ij}$ represents a real $4\times4$ rotation in the ($i,j$) plane and
the matrix A is the product of the three matrices ($R_{23},  R_{24}, R_{34}$).%
%%%%%%%%%%%%%%%%%%%%%%%%%%%%%%%%%%%%%%%%%%%%%%%%%
\footnote{The order of the three matrices in the product is irrelevant as we are interested only in the electron
neutrino mixing. We have dropped the dependence on the (three) CP-violating phases as they are
unobservable in reactor setups.}
%%%%%%%%%%%%%%%%%%%%%%%%%%%%%%%%%%%%%%%%%%%%%%%%%
This parameterization leads to the following simple expressions for the matrix elements involving 
the electron neutrino
% .................................................................................................................
\begin{align}
    \label{eq:Ue1} U_{e1} 
     &= \cos\theta_{14}  \cos\theta_{13} \cos\theta_{12}\,, \\\
\label{eq:Ue2} U_{e2}
        &=  \cos\theta_{14}  \cos\theta_{13} \sin\theta_{12}\,,    \\
\label{eq:Ue3} U_{e3} 
        & = \cos\theta_{14}  \sin\theta_{13}\,,  \\
\label{eq:Ue4}  U_{e4}
        &= \sin\theta_{14} \,,
\end{align}
% .................................................................................................................
and as a consequence the approximate survival probability  in Eq.~(\ref{eq:degouvea}) takes the form
%%%%%%%%%%%%%%%%%%%%%%%%%%%%%%%%%%%%%%%%%%%%%%%%%%%%%%%%%%
\begin{equation}
P_{ee}\simeq 1 -\sin^22\theta_{13}\sin^2\left(\frac{\Delta m^2_{13}L}{4E_{\nu}}\right)-\sin^22\theta_{14}\sin^2\left(\frac{\Delta m^2_{14}L}{4E_{\nu}}\right)\,.
\label{eq:pee_approx}
\end{equation}
%%%%%%%%%%%%%%%%%%%%%%%%%%%%%%%%%%%%%%%%%%%%%%%%%%%%%%%%%%
In the limit $\theta_{14}\to 0$ one recovers the well-known expression for $P_{ee}$ at $\theta_{13}$-driven reactor neutrino experiments.%
%%%%%%%%%%%%%%%%%%%%%%%%%%%%%%%%%%%%%%%%%%%%%%%%%%%%%%%%%%
\footnote{We stress that in our codes we have incorporated the general expression of the survival probability given in Eq.~(\ref{eq:general}), 
limiting the usage of the approximated expression in Eq.~(\ref{eq:pee_approx}) to qualitative discussions.}

In order to qualitatively understand the effect of sterile neutrinos in reactor neutrino setups it useful to express the phase entering the oscillator factor of the $\theta_{14}$-driven term in Eq.~(\ref{eq:pee_approx}) as follows
%%%%%%%%%%%%%%%%%%%%%%%%%%%%%%%%%%%%%%%%%%%%%%%%%%%%%%%
\begin{equation}
\frac{\Delta m^2_{14}L}{4E_{\nu}}\simeq1.267 \left( \frac{\Delta m^2_{14}}{10^{-2}~\rm eV^2}\right)\left(\frac{L}{400~\rm m}\right)\left(\frac{4~\rm MeV}{E_{\nu}}\right).
\label{eq:phase}
\end{equation}
%%%%%%%%%%%%%%%%%%%%%%%%%%%%%%%%%%%%%%%%%%%%%%%%%%%%%%%
With the help of Eq.~(\ref{eq:phase}) we can make the following observations. I)  For values in the interval  $\Delta m^2_{14} \in [10^{-3} - 10^{-1}]$~eV$^2$ sterile neutrino oscillations can affect $P_{ee}$ in the near and far detectors in distinct ways. Therefore, one expects that the near-versus-far comparison is affected by the presence of VLS$\nu$ oscillations.
II) For values of $\Delta m^2_{14} \lesssim 10^{-3}$~eV$^2$ the sensitivity quickly decreases as the phase in Eq.~(\ref{eq:phase}) becomes small both at the near and far site.  III) For values of $\Delta m^2_{14} \gtrsim 10^{-1}$~eV$^2$ the 
phase becomes large and the oscillations get averaged due to the integration over the energy both
 at the near and far detectors of Daya Bay and RENO, which are positioned at distances of at least few hundreds meters from the reactor cores. As a consequence, the effect of the oscillations cancel out in the near/far ratio of their total rates, which becomes insensitive to sterile neutrino oscillations. IV) In the special case of Double Chooz, the role of the near detector is played by Bugey-4, which is 
 a very short baseline ($\sim 15~$m) experiment.  For this reason, in principle one could obtain information also on values
 of $\Delta m^2_{14}$ lying outside (above) the range under investigation. However, as in this paper we are not 
 interested in such values,  we will not consider such a potentiality.
 
\section{Production and detection of reactor antineutrinos}

\subsection{Production}

Nuclear reactors release electron antineutrinos produced in the beta decays of the fission products of $^{235}$U, $^{238}$U, $^{239}$Pu, and $^{241}$Pu.   
In general, in a reactor experiment, a given detector (labelled by the index $d$) 
will receive neutrinos from a number of reactors (each labelled by the index $r$) having average 
thermal powers $P^{\mathrm {th}}_r$
and located at distances $L_{r,d}$ from the detector.
The time-averaged differential neutrino flux (number of neutrinos  per unit of time, area, and energy) 
at such a detector due to the $r$-th reactor is  given by
%...................................................................
\begin{equation}
\label{nuflux}
\left(\frac{d\phi}{dE_\nu}\right)_{r,d} \simeq \frac{P^\mathrm{th}_r}{4\pi L_{r, d}^2}
\frac{\sum_ k \langle \alpha_k^r\rangle  S_k(E_\nu)}{\sum_k \langle \alpha_k^r\rangle \langle E_k\rangle}\,,
\end{equation}
%.....................................................................
where $\langle \alpha_k^r \rangle$ is the time-averaged fractional contribution of the $k$-th fissile nuclei to the total number of fissions occurring in the $r$-th reactor, $\langle E_k \rangle$ is the corresponding average fission energy and $S_k(E_\nu)$  is the neutrino spectrum per fission
of the $k$-th branch.  For equilibrium pressurized light water reactors, the fuel composition is very similar. Therefore, 
lacking a detailed information on the fractions $\langle \alpha_k^r\rangle $ we used for all the reactors a common fuel
composition $\langle \alpha_k\rangle \equiv \langle \alpha_k^r\rangle $ equal to that of Bugey~\cite{Declais:1994ma}
%...................................................................
\begin{equation}
^{235}\mathrm{U}:\,
^{238}\mathrm{U}:\,
^{239}\mathrm{Pu}:\,
^{241}\mathrm{Pu}=0.538:0.078:0.328:0.056\,.
\end{equation}
%.....................................................................
We have checked that this approximation induces negligible inaccuracies in the estimate
of the ratios of the number of events expected with and without oscillations (given in Eq.~(\ref{eq:ratio})),
which will be relevant for our analysis. Concerning the average fission energies we assume for
their values~\cite{Kopeikin:2004cn} ($201.92,205.52, 209.99, 213.60)\,$MeV, as done by all the three experimental collaborations.
We parametrize the $k$-th spectral component as an exponential of a polynomial function~\cite{Vogel:1989iv}
%%%%%%%%%%%%%%%%%%%%%%%%%%%%%%%%%%%%%%%%%%%%%%
\begin{equation}
 S_k(E_\nu) =  \exp \left[ \sum_{j=1}^6 a_j^kE_{\nu}^{(j-1)}  \right],
 \label{eq:spectrum}
 \end{equation}
%%%%%%%%%%%%%%%%%%%%%%%%%%%%%%%%%%%%%%%%%%%%%%
taking the $a_j^k$ coefficients from~\cite{Huber:2011wv}. In the presence of flavor oscillations,
Eq.~(\ref{nuflux}) must be multiplied by the neutrino survival probability $P_{ee}(E_\nu,L_{r,d})$.

\subsection{Detection}

Reactor antineutrinos are observed trough the inverse
beta decay (IBD) reaction 
\mbox{$\bar{\nu}_e + p \rightarrow e^+ + n$}.
While the positron signal is promptly detected, the 
neutron's one is seen after a mean time of about 30 $\mu$s from its
capture in the gadolinium-doped target.
The measurement of the prompt energy of the positron allows the determination of the
antineutrino energy. The energy  $E$ deposited by the positron
(including annihilation) is related to antineutrino energy $E_\nu$ by
%%%%%%%%%%%%%%%%%%%%%%%%%%%%%%%%%%%%%%%%
\begin{equation}
E = E_{\nu} - T_n - 0.782~{\mathrm {MeV}}\,,
\end{equation}
%%%%%%%%%%%%%%%%%%%%%%%%%%%%%%%%%%%%%%%%
where $T_n$ denotes the average neutron recoil energy, 
which is small compared to $E_{{\nu}}$.
Given the differential neutrino flux in Eq.~(\ref{nuflux}), the 
time-averaged energy spectrum of events (number of 
expected events per
unit of prompt positron energy $E$)  at the $d$-th detector due to the $r$-th reactor is given by 
%...................................................................
\begin{equation}
\label{dNdE}
\left(\frac{dN}{dE}\right)_{r,d}=\varepsilon_d \, n_d\, M_d\,\Delta t_d
\int dE_\nu\, \left(\frac{d\phi}{d E_\nu}\right)_{r,d}\int dE'\, \frac{d\sigma(E_\nu,E')}{dE'}\,r_d(E,E')\ ,
\end{equation}
%.....................................................................
where $\varepsilon_d$ is the detector overall efficiency after all cuts, 
$n_d$ is the target density, $M_d$ is its mass,
$\Delta t_d$ is the live-time,
$r_d(E,E')$ is the energy resolution function, and
$\sigma$ is the IBD cross section.
For the energy resolution function we have assumed a gaussian shape 
having width reported in the published papers:   [$7.6/\sqrt{E'({\rm MeV)}}$] \% 
for Double Chooz,  [$5.9/\sqrt{E'({\rm MeV)}} + 1.1$]\% for RENO and 
[$7.5/\sqrt{E'({\rm MeV)}} + 0.9$]\% for Daya Bay. The differential cross section can be estimated as
%...................................................................
\begin{equation}
\frac{d\sigma(E_\nu,E')}{dE'}\simeq \sigma(E_\nu)\,
\delta(E_\nu-E'-0.782~\mathrm{MeV})\ ,
\end{equation}
%.....................................................................
with $\sigma(E_\nu)$ taken from~\cite{Vogel:1999zy}. 
The total number of events theoretically expected in the presence of oscillations 
is obtained by integrating Eq.~(\ref{dNdE}) over the prompt energy
%.................................................................
\begin{equation}
N_{r,d}^{\rm the} = \int_0^{\infty} \left(\frac{dN}{dE}\right)_{r,d}dE\,.
\end{equation}
%.....................................................................
Finally, the oscillated-over-non-oscillated ratio of the events  
 is given by
%...................................................................
\begin{equation}
R_{r,d} = \frac{N_{r,d}^{\rm{the}}}{N^0_{r,d}} \equiv \langle P_{ee}\rangle_{r,d}\,,
\label{eq:ratio}
\end{equation}
%.....................................................................
where $N^0_{r,d}$ is calculated taking $P_{ee}(E_\nu, L_{r,d}) = 1$. Obviously,
the ratio in Eq.~(\ref{eq:ratio}) is independent of the normalization parameters of the
detector ($\varepsilon_d, n_d, M_d, \Delta t_d$) and of the reactor power $P^{\rm th}_r$,
and it is also independent of the geometrical factor $1/L_{r,d}^2$. The only dependence
on the distance $L_{r,d}$ is that one encoded in the survival probability. In fact, the ratio in Eq.~(\ref{eq:ratio}) is
nothing else than the survival probability convoluted with the neutrino energy spectrum, 
the IBD cross-section and the detector energy resolution. In Eq.~(\ref{eq:ratio}) we have indicated
such an equivalence by introducing the effective survival probability $\langle P_{ee}\rangle_{r,d} \equiv \langle P_{ee} (L_{r,d})\rangle$.

The total number of events expected at the detector $d$ will be the sum of the partial
contributions arising from all reactors of the experiment
%...................................................................
\begin{equation}
N_d^{\rm the} = \sum_r N^{\rm the}_{d, r} = \sum_r N_{r,d}^0  \langle P_{ee}\rangle_{r,d} \equiv 
N_{d}^0 \sum_r  \omega_r^d  \langle P_{ee}\rangle_{r,d} \,,
\label{eq:Nd}
\end{equation}
%.....................................................................
where we have introduced the total number of non-oscillated events $N_d^0$ expected at the $d$-th detector and 
the fractions $\omega_r^d$ of such events induced by the $r$-th reactor. These last ones only depend on baselines and thermal powers, and
can be expressed as 
%%%%%%%%%%%%%%%%%%%%%%%%%%%%%%%%%%%%%%%%%%%%%%%%%%%%%%
\begin{equation}
 \omega_r^d = \frac{w_r/L_{r, d}^2}{\sum_r(w_r/L_{r, d}^2)}
 ~~{\rm with}~~
 w_r =\frac{P_r^{\rm th}}{\sum_r P_r^{\rm th}}\,,
 \label{eq:omegas}
\end{equation}
%%%%%%%%%%%%%%%%%%%%%%%%%%%%%%%%%%%%%%%%%%%%%%%%%%%
where $w_r$ is the relative weight of the $r$-th reactor to the total power emitted
by all reactors. All the relevant information on the detector normalization parameters 
($\varepsilon_d, n_d, M_d, \Delta t_d$) is encoded in the theoretical
non-oscillated number of events $N_{d}^0$ in Eq.~(\ref{eq:Nd}), while the fractions 
$\omega_r^d$ only depend on the reactor powers and baselines.
For all the experiments we will take (or derive) the 
numbers $N_d^0$ and the fractions $\omega_r^d$ from the published papers.

Finally, for later use, we introduce the reactor-flux-weighted baseline of the $d$-th detector
%%%%%%%%%%%%%%%%%%%%%%%%%%%%%%%%%%%%%%%%%%%%%%%%%%%%%%
\begin{equation}
 L_d = \sum_r f_{r}^{d} L_{r,d}\,,
 \label{eq:wei_baselines}
\end{equation}
%%%%%%%%%%%%%%%%%%%%%%%%%%%%%%%%%%%%%%%%%%%%%%%%%%%
where each baseline has an effective fractional weight $f_r^d$ proportional to the geometrical factor $1/L_{r,d}^2$
%%%%%%%%%%%%%%%%%%%%%%%%%%%%%%%%%%%%%%%%%%%%%%%%%%%%%%
\begin{equation}
f_r^d =  \frac{1/L^2_{r,d}}{\sum_r 1/L_{r,d}^2}\,.
 \label{eq:frac}
\end{equation}
%%%%%%%%%%%%%%%%%%%%%%%%%%%%%%%%%%%%%%%%%%%%%%%%%%%
The reactor-flux-weighted  
baselines will be helpful in the interpretation of the results obtained in multi-reactor experiments like
RENO and Daya Bay.

\section{Statistical analysis of reactor neutrino data}

\subsection{Double Chooz}

The Double Chooz (DC) experiment makes use of the two Chooz reactors each having
a thermal power of  $4.25~{\rm GW}$. Currently, the experiment is operating only with the far detector, 
since the near detector is not complete yet.
This renders problematic the extraction of the parameter $\theta_{13}$ as its estimate 
would strongly depend on the theoretical absolute reactor neutrino fluxes.
%are subjected to and have been recently recalculated to be larger than before.
To mitigate this problem the collaboration used as an ``anchor'' the total rate measurement
made by the short baseline experiment Bugey-4~\cite{Declais:1994ma}, which provides a rather precise measurement
(with a 1.4\% error) of the reactor antineutrino flux. In this way, the estimate of $\theta_{13}$ 
is independent of the normalization of the reactor fluxes and is also 
independent of the sterile neutrino oscillations possibly occurring at short baselines.%
%%%%%%%%%%%%%%%%%%%%%%%%%%%%%%%%%%%%%%%%%%%%%%%
\footnote{This is true only for $\Delta m_{14}^2 \gtrsim 3~\mathrm {eV}^2$, for which the sterile neutrino
oscillations get completely averaged at the Bugey-4 site (see the discussion in~\cite{Giunti:2011vc}).}
%%%%%%%%%%%%%%%%%%%%%%%%%%%%%%%%%%%%%%%%%%%%%%%
The use of Bugey-4 as an anchor point is practically equivalent to consider it as a near-site detector of the experiment. 
Due to the very short baseline (15~m), when working in a 3-flavor framework the expected rate of Bugey-4 can 
be considered independent of $\theta_{13}$. Differently, when considering a 4-flavor scheme, 
one has to take into account that it can depend on the mass-mixing parameters, as we have done
in our simulations.

While  Double Chooz has performed a rate plus shape analysis, from outside the collaboration it is not possible
to reproduce the spectral analysis since detailed information on the spectrum, in particular the bin-to-bin 
correlated errors, is missing. So, we restricted the analysis to the total rate information. We define the   
Double Chooz $\chi^2$ estimator as follows
%%%%%%%%%%%%%%%%%%%%%%%%%%%%%%%%%%%%%%%%%%%%%%%
\begin{equation}
 \chi^2_{\rm Double~ Chooz}  =  
\frac{\left(\frac{r^{\rm obs}}{r^{\rm the}}- 1\right)^2} {\sigma^2}\,,   
\label{eq:chi2_DoubleChooz}
\end{equation}
%%%%%%%%%%%%%%%%%%%%%%%%%%%%%%%%%%%%%%%%%%%%%%%
 where $r^{\rm obs}$ ($r^{\rm the})$ is the ratio of the observed (theoretical) total fission cross-section
 of DC over that of Bugey-4. We have taken the 
relative error on the ratio $\sigma  = 0.028$ as estimated from~\cite{Abe:2012tg}. 

\subsection{RENO}

%%%%%%%%%%%%%%%%%%%%%%%%%%%%%%%%%%%%%%%%%%%%%%%%%%%%%%%%
\begin{table}[t]
 \caption{ \label{table:RENO_baselines}
For each of the two RENO detectors (ND, FD), the first row reports the baselines, which have been estimated from the 
relative flux contributions. The second row reports
the fractional neutrino flux contributions referring to the no-oscillation case and calculated for
the following reactor powers (R1, R2 relative power = 95.6\%, R3-R6 relative power = 100\%)~\cite{RENO_private}. The fractional contributions account for the burn-up corrections of the six reactors~\cite{RENO_private}.
}
\begin{ruledtabular}
  \begin{tabular}{lrrrrrr}
      & R1     & R2     & R3     & R4     & R5     & R6 \\
   \colrule
   ND baseline [m]&  660.6 &  445.0 &  301.3 &  339.0 &  519.9 &  746.3 \\
   ND contribution  & 0.0674 & 0.1488 & 0.3612 & 0.2567 & 0.1078 & 0.0581 \\
   \hline
    FD baseline [m] & 1564.2 & 1461.0 & 1397.8 & 1380.0 & 1409.3 & 1483.0 \\
   FD contribution  & 0.1372 & 0.1568 & 0.1917 & 0.1771 & 0.1688 &  0.1684\\
   \end{tabular}
 \end{ruledtabular}
\end{table}
%%%%%%%%%%%%%%%%%%%%%%%%%%%%%%%%%%%%%%%%%%%%%%%%%%%%%%%%

%%%%%%%%%%%%%%%%%%%%%%%%%%%%%%%%%%%%%%%%%%%%%%%%%%%%%%%%
\begin{table}[t]
 \caption{  \label{table:RENO_parameters}
Fitting parameters used for the RENO data analysis taken from~\cite{RENO_NANPino}.
 }
\begin{ruledtabular}
  \begin{tabular}{lrr}
      & ND     & FD  \\
   \colrule
   $N_d^{0}$  & 275535.61        & 30400.25         \\
    $N_d^{\rm obs}$    & 272229.74        & 28240.65         \\
   $\sigma_d^\xi$       & 0.002            & 0.002            \\
   $\sigma_d^b$		& 786.03		   & 241.61 
     \end{tabular}
 \end{ruledtabular}
\end{table}
%%%%%%%%%%%%%%%%%%%%%%%%%%%%%%%%%%%%%%%%%%%%%%%%%%%%%%%%

The RENO experiment makes use of two identical detectors, one near (ND) and one far (FD), detecting neutrinos produced in six nuclear reactors (R1-R6). The distances of each detector reactor pair are not published but, given the simple geometry of the setup, they can be deduced from the fractional (non-oscillated) flux contributions $\omega_r^d$ of each reactor to each detector. We used the fractions that we obtained from the collaboration~\cite{RENO_private} reported in 
Table~\ref{table:RENO_baselines}, and which refer to a relative power of 95.6\% for the reactors R1 
and R2, and 100\% for the reactors R3-R6. For the weighted baselines of the near and far detectors we find respectively $408.5$\,m and $1444.1$\,m 
in excellent agreement with those provided by the collaboration ($408.6$\,m and $1444.0$\,m).  
For our analysis we made use of the data taken in the period (Aug. 2011 - Oct. 2012) first presented at the Neutrino Telescopes 2013 conference~\cite{RENO_neutel}, which refer to a longer live-time (369.034 days for near detector, 402.693 days for the far detector) with respect to
that used for the analysis presented in the published paper~\cite{Ahn:2012nd}.  Following the collaboration, we construct the $\chi^2$ with pull terms accounting for the correlation of the systematic errors, as follows
%%%%%%%%%%%%%%%%%%%%%%%%%%%%%%%%%%%%%%%%%%%%%%%%%%%%%%% 
\begin{equation}
%\begin{split}
\chi^2_{\rm RENO} = \\ 
\sum_{d}^2 \frac{\left[ N^{\rm obs}_d + b_d - (1 + a + \xi_d) \sum_{r}^6 (1+\alpha_r) N^{\rm the}_ {r,d}\right]^2}{N^{\rm obs}_ d} 
\\
 +\sum_{d}^2 \left( \frac{\xi_d^2}{(\sigma_d^{\xi})^2} + \frac{b_d^2}{(\sigma_d^b)^2}\right)
 + \sum_{r}^6 \frac{\alpha_r^2}{\sigma_r^2},
%\end{split}
\label{eq:chi2_RENO}
\end{equation}
%%%%%%%%%%%%%%%%%%%%%%%%%%%%%%%%%%%%%%%%%%%%%%%%%%%%%%%
where $d$ is an index denoting the near or the far detector, the $r$ index
corresponds to reactors 1 through 6, $N^{\rm obs}_d$ is the number of observed IBD candidates in each detector after background subtraction and $N^{\rm the}_{r,d}$ is the number of expected neutrino events due to the $r$-th reactor to the $d$-th detector. 
The uncorrelated reactor uncertainty (common to all reactors) is $\sigma_r$ (amounting to 0.9\%), $\sigma_d^{\xi}$ is the uncorrelated detection uncertainty, and $\sigma_d^b$ is the background uncertainty. Apart from the fractions $\omega_r^d$ that we obtained directly from the RENO collaboration~\cite{RENO_private}, all the other parameters needed to build Eq.~(\ref{eq:chi2_RENO}) were taken from~\cite{RENO_NANPino}.  
The parameters related to the two detectors entering Eq.~(\ref{eq:chi2_RENO}) are reported in Table~\ref{table:RENO_parameters}.
In each point of the mass-mixing parameter space, the 
$\chi^2$ was minimized with respect to the 10 ``pulls''  [$\alpha_r$(6),$~\xi_d$(2),$~b_d$(2)] and the (free) normalization parameter $a$, which takes into account our ignorance on the absolute neutrino fluxes.

\subsection{Daya Bay}

The Daya Bay experiment consists of three experimental halls (EH1-EH3), each containing one or more antineutrino detectors (AD).
The ADs receive neutrinos from six reactors grouped in three power stations each consisting of two reactor cores.
 The reactor-flux-weighted baselines of the three halls are given by  $470$\,m (EH1), $576$\,m (EH2) and $1648$\,m (EH3),
so the experiment has one far and two near sites.
All reactors are functionally identical pressurized water reactors with maximum thermal power of  2.9~GW~\cite{An:2012bu}. 

We define the $\chi^2$ as in the Daya Bay publication~\cite{An:2012bu}, which after a slight change of notation made for homogeneity with the treatment of the RENO data, can be expressed as%
%%%%%%%%%%%%%%%%%%%%%%%%%%%%%%%%%%%%%%%%%%%%%%%%%%%%%%%
\begin{equation}
 \chi^2_{\rm Daya\, Bay} =   
\sum_{d}^6 \frac{\left[ N_d^{\rm obs}  + b_d - N_d^{\rm the} \left( 1+a + \sum_r^{6}  \omega_r^d \alpha_r + \xi_d \right) \right]^2}{N_d^{obs} + b_d} 
  + \sum_d^6 \left( \frac{\xi_d^2}{(\sigma_d^\xi)^2} + \frac{b_d^2}{(\sigma_d^b)^2} \right)
 + \sum_r^{6} \frac{\alpha_r^2}{\sigma_r^2} \label{eq:chi2_DayaBay}\,,
\end{equation}
%%%%%%%%%%%%%%%%%%%%%%%%%%%%%%%%%%%%%%%%%%%%%%%%%%%%%%%
where the quantities with identical naming have the same meaning as in RENO. The relevant parameters
for the six detectors (AD1-AD6) are provided in Table~\ref{table:DayaBay_Parameters}.
The relative error on the reactor normalization is $\sigma_r = 0.8\%$. Compared to RENO, in this case the number of pulls is larger (we have 18 pulls in total
plus the normalization parameter $a$) since there are six detectors. 
While the baselines $L_{r, d}$ are provided in~\cite{An:2012bu}, no detailed information
is given on the reactor powers. Lacking a more detailed information, we assumed $w_r = 1/6$ for the
calculation of the fractions $\omega_r^d$ defined in Eq.~(\ref{eq:omegas}), 
since all reactors have identical nominal thermal power.

%%%%%%%%%%%%%%%%%%%%%%%%%%%%%%%%%%%%%%%%%%%%%%%%%%%
\begin{widetext}
\begin{center}
\begin{table}[t]
 \caption{  \label{table:DayaBay_Parameters} Fitting parameters used for the Daya Bay analysis taken from~\cite{An:2012bu}.}
 \begin{ruledtabular}
  \begin{tabular}{crrrrrr}
       & AD1 & AD2   & AD3 & AD4 & AD5 & AD6 \\ 
   \colrule
    $N_{d}^{0}$          & 68613   & 69595 & 66402 & 9922.9 & 9940.2 & 9837.7 \\
   $N_d^{obs}$              & 67722.45 & 68333.73 & 65367.12 & 9358.50 & 9240.84 & 9037.32 \\
   $\sigma_d^\xi$         & 0.002 &  0.002 & 0.002 & 0.002 & 0.002 & 0.002 \\
  $\sigma_d^b$         & 157.03 & 156.46 & 124.53 & 28.75 & 28.74 & 28.55 
  \end{tabular}
 \end{ruledtabular}
\end{table}
\end{center}
\end{widetext}
%%%%%%%%%%%%%%%%%%%%%%%%%%%%%%%%%%%%%%%%%%%%%%%%%%%

\subsection{Combination}

The global $\chi^2$ of the combination of the three experiments is simply obtained by summing up the
single contributions of each single experiment as follows
%%%%%%%%%%%%%%%%%%%%%%%%%%%%%%%%%%%%%%%%%%%%
\begin{equation}
\chi^2_{\rm TOT} = \chi^2_{\rm Double\, Chooz} + \chi^2_{\rm RENO} + \chi^2_{\rm Daya\, Bay}
\end{equation}
%%%%%%%%%%%%%%%%%%%%%%%%%%%%%%%%%%%%%%%%%%%%
where possible correlations among different experiments have been neglected. This is well justified
for all the errors appearing in the Eqs.~(\ref{eq:chi2_DoubleChooz}-\ref{eq:chi2_DayaBay})
with the exception of the free normalization correction $a$, which in principle should be treated as a 
common parameter in Daya-Bay and RENO. 
However, it must be observed that the expected absolute non-oscillated number of events 
published by both experiments are not obtained by an ab initio calculation
as that one sketched in Sec.~III. Instead, they are numbers obtained {\em after} correction with the (best fit) 
value of a normalization parameter derived by the $\chi^2$ minimization in a 3-flavor scheme.
This is confirmed by our simulations, which, at the best fit point of the 3-flavor analysis, provide very small
departures from  zero of the normalization parameter $a$.
When (hopefully) Daya Bay and RENO will provide the non-oscillated number of events expected at each
detector as derived by an ab initio calculation, it will be possible (and correct) to use a common
normalization parameter for both experiments. In such a case, their combined allowed/excluded mass-mixing
parameters regions are expected to be slightly more restrictive with respect to those ones obtained in our analysis.

\section{Three flavor analysis}

Before discussing the results of the full 4-flavor analysis we deem it useful to 
consider in some detail the simple 3-flavor case, which is obtained in the limit of $\theta_{14} = 0$.
In this case, the electron neutrino survival probability, neglecting solar neutrino terms,
depends only on the atmospheric mass-squared splitting $|\Delta m^2_{13}|$ and from $\theta_{13}$.
For the sake of precision in our analysis we include also the solar terms, fixing
the solar parameters at their best fit values as obtained in the global analysis performed in~\cite{Fogli:2012ua}.

%%%%%%%%%%%%%%%%%%%%%%%%%%%%%%%%%%%%%%%%%%%
\begin{figure*}[t!]
\vspace*{-7.5cm}
\hspace*{1.00cm}
\includegraphics[width=20.0 cm]{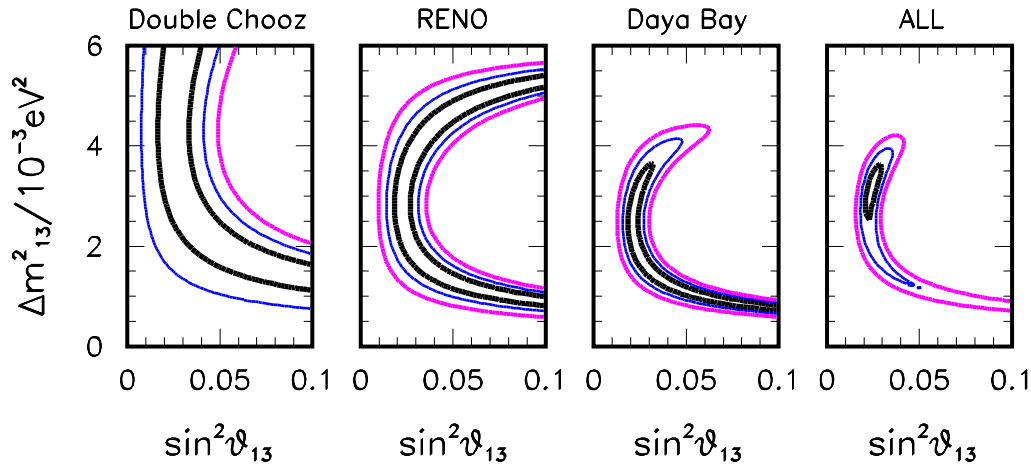}
\vspace*{-6.8cm}
\caption{Results obtained in the 3-flavor case ($\theta_{14} = 0$).
Regions allowed by the three reactor experiments and by their combination.  
The contours refer to $\Delta \chi^2 =1,4,9$.
\label{fig1}}
\end{figure*}  
%%%%%%%%%%%%%%%%%%%%%%%%%%%%%%%%%%%%%%%%%%%%%

Before commenting the 3-flavor analysis results it is opportune to compare
our findings for the estimate of $\theta_{13}$  with those published by each single experiment.
For this purpose we fix the value of the atmospheric mass-squared splitting at the MINOS best fit~\cite{Adamson:2011ig} $|\Delta m^2_{13}| = 2.32 \times 10^{-3}\,\mathrm{eV}^2$, as done by all the three experiments.
We obtain for Double Chooz, $\sin^2 2\theta_{13} = 0.154 \pm 0.053$  ($\theta_{13} = 0$ excluded 
at 2.9 $\sigma$) in good agreement
with the rate-only analysis made by the collaboration, which gives $\sin^2 2\theta_{13} = 0.170 
\pm 0.052$ ($\theta_{13} = 0$ excluded at 3.3 $\sigma$). For RENO we obtain $\sin^2 2\theta_{13} = 0.098\pm 0.019$ ($\theta_{13} = 0$ excluded 
at 5.2 $\sigma$), in agreement with the collaboration findings $\sin^2 2\theta_{13} = 0.100\pm 0.018$
 ($\theta_{13} = 0$ excluded  at 5.6 $\sigma$). For Daya Bay 
we obtain $\sin^2 2\theta_{13} = 0.086 \pm 0.012$ ($\theta_{13} = 0$ excluded 
at 7.4 $\sigma$), again in excellent agreement with the
collaboration result $\sin^2 2\theta_{13} = 0.089\pm 0.011$  ($\theta_{13} = 0$ excluded 
at 7.7 $\sigma$). Finally, the estimate obtained
from the combination of the three experiments is $\sin^2 2\theta_{13} = 0.090\pm 0.009$,
representing an evidence of non-zero $\theta_{13}$ at the $\sim 10$ sigma level.

Reassured by these checks, which make us confident on the accuracy of our analysis, we show 
in Fig.~1 the 2-dimensional regions determined by the three experiments taken separately
(first three panels) and by their combination (fourth panel).
The curves represent contours at the $\Delta \chi^2 =1, 4, 9$ levels. With this choice,
one can easily deduce the allowed ranges for the single parameters at the $1,2,3\, \sigma$ levels,
by just projecting the 2-dimensional region onto the corresponding axis. 
It must be noted that all the three collaborations, Double-Chooz, RENO and Daya Bay, consider
the estimate of $|\Delta m^2_{13}|$ as an external input, since this parameter is precisely determined by MINOS.
In our analysis we can leave this parameter free to vary. This allows
us to check that the estimates of $|\Delta m^2_{13}|$ obtained by the three experiments 
are mutually consistent and that their global estimate is consistent with that obtained from 
the accelerators, which make use of  a different measurement  technique based on 
 $\nu_\mu \to \nu_\mu$ disappearance.

From the last panel in Fig.~1 we find that the global reactor estimate of the atmospheric 
splitting is $|\Delta m^2_{13}| = 3.2_{-2.0}^{+0.8}\times 10^{-3}\,\mathrm{eV}^2$ (2 $\sigma$ level), which
is consistent with the determination obtained from MINOS $|\Delta m^2_{13}| = 2.32_{-0.16}^{+0.24}\times 10^{-3}\,\mathrm{eV}^2$ (2 $\sigma$ level), although it is much less precise.% 
%%%%%%%%%%%%%%%%%%%%%%%%%%%%%%%%%%%%%%%%%%%%%%%%
\footnote{The region in the fourth panel of Fig.~1 is in agreement with the analogous one
presented in~\cite{GonzalezGarcia:2012sz,Bezerra:2013mua},
where similar 3-flavor analyses have been performed.}
%%%%%%%%%%%%%%%%%%%%%%%%%%%%%%%%%%%%%%%%%%%%%%%%
We observe a slight mismatch between the determination of both parameters $\theta_{13}$ and $|\Delta m^2_{13}|$ 
made by Double Chooz and those made by the other two experiments RENO and Daya Bay, which is driven
by the bigger rate suppression observed by Double Chooz. The comparison of the first panel with the second
and third ones, shows that Double Chooz is complementary to RENO and Daya-Bay in the determination
of  $|\Delta m^2_{13}|$. As already observed in~\cite{Bezerra:2013mua}, this behavior 
is due to the fact that at the Double Chooz far site baseline ($\sim 1~$km) the survival probability 
has maximal (negative) slope, while at RENO ($\sim 1.4~$km) and Daya-Bay
($\sim 1.6~$km) it is close to its minimum.  For this reason, the more precise measurement  expected
by Double Chooz with the near detector will be very important to improve the global accuracy
of the (reactor) estimate of  $|\Delta m^2_{13}|$. 

To this regard we remind that the comparison of the measurements of the atmospheric mass-squared 
splitting made at reactors (based on $\nu_e \to \nu_e$ disappearance) with that performed at the accelerators
(based on $\nu_\mu\to \nu_\mu$ disappearance) could be used in principle to the determine 
the neutrino mass hierarchy~\cite{Nunokawa:2005nx}. However, at the moment we are very far from 
realizing such a possibility, as it requires an accuracy at the sub-percent level in both measurements~\cite{Nunokawa:2005nx}.

\section{Four flavor analysis}

In the 4-flavor analysis we fix the atmospheric mass-squared splitting at the best fit
value obtained by the global analysis performed in~\cite{Fogli:2012ua}. We have checked
that the results  show negligible differences varying this
parameter in the interval currently allowed. As done in the 3-flavor case we also fix the solar parameters
 at their best fit values. Therefore, the analysis depends on three parameters:
the new mass-splitting $\Delta m^2_{14}$ and the two mixing angles $\theta_{13}$ and
$\theta_{14}$.  We allow both mixing angles to vary in the range $\theta_{ij}\leq \pi/4$,
ignoring the bigger uninteresting values lying in the ``dark octants'' corresponding to $\theta_{ij}> \pi/4$.
We have checked that the results of the analysis show only a negligible dependence on the sign
of $\Delta m^2_{14}$ as expected from the qualitative discussion made in Sec.~II (see also~\cite{deGouvea:2008qk}).
Therefore, for definiteness we show the results only for $\Delta m^2_{14}>0$.
In consideration of the upper limit established in~\cite{Palazzo:2013me} 
(see also~\cite{Palazzo:2011rj,Palazzo:2012yf,Giunti:2012tn}) with the combination of solar and KamLAND data 
[$\sin^2\theta_{14}<0.08$ (95\%\, C.L. 1 d.o.f.)],  
which is independent on $\Delta m^2_{14}$ as far as it is much bigger than the solar
mass-squared splitting (a condition fulfilled for the range of values explored in this paper), we  
display the results of our analysis in the phenomenologically interesting region of $\sin^2\theta_{14}<0.1$.
For a better clarity we first discuss the results obtained for a fixed value of $\theta_{13}$ chosen at the best
fit point of the 3-flavor analysis. Then we discuss  the general case in which $\theta_{13}$ 
is treated as a free parameter.  

%%%%%%%%%%%%%%%%%%%%%%%%%%%%%%%%%%%%%%%%%%%
\begin{figure*}[t!]
\vspace*{-7.5cm}
\hspace*{1.00cm}
\includegraphics[width=20.0 cm]{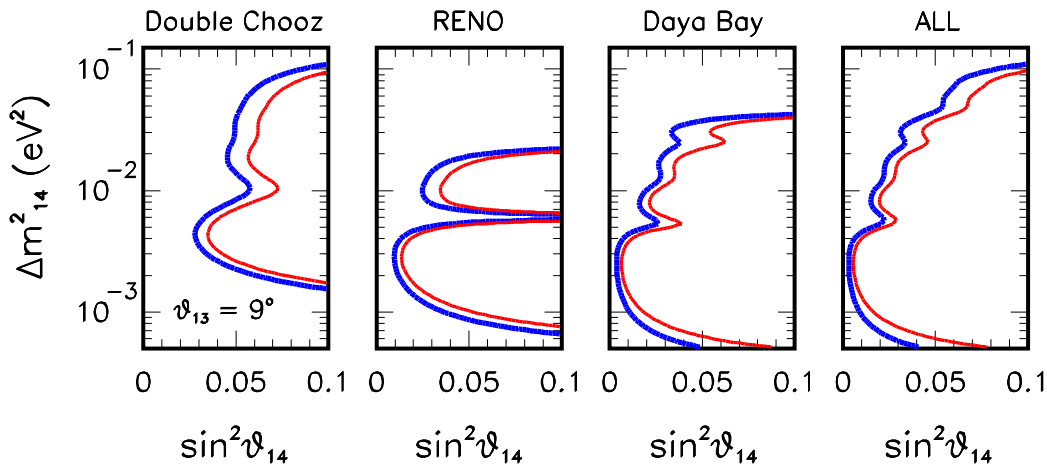}
\vspace*{-6.8cm}
\caption{Results obtained from the 4-flavor analysis for a fixed value of $\theta_{13} = 9^o$. The contours refer to 2 d.o.f. 90\% C.L. (blue thick line) and 99\% (red thin line).
\label{fig2}}
\end{figure*}  
%%%%%%%%%%%%%%%%%%%%%%%%%%%%%%%%%%%%%%%%%%%%%

\subsection{The  case of fixed $\theta_{13}$} 

Figure~2 shows the results of the 4-flavor analysis for the particular case of $\theta_{13}$
{\em fixed} at its best fit value as obtained in the 3-flavor analysis ($\theta_{13} \simeq 9^o$).
Therefore, the contours represent a section of the 3-dimensional space spanned by the three parameters. 
This particular case will help in the interpretation of the most general case, in which the parameter $\theta_{13}$
will be left free to vary and marginalized away when taking the 2-dimensional projection in the
plane of the two displayed parameters. We plot the confidence levels for 2 d.o.f.  at  90\% and 99\% as 
usually done in the literature, in order to facilitate the comparison of our results with those already
existing. 

In Fig.~2, the first panel represents the regions excluded 
by Double Chooz. Around $\Delta m^2_{14} \sim 4\times 10^{-3}$\,eV$^2$, where the exclusion
regions are more restrictive, the 4-flavor (and 3-flavor) effects at the near site (Bugey-4), which
has a baseline of only 15\,m are completely negligible. Therefore, in this range
of parameters Bugey-4  provides a measurement of the non-oscillated flux. The comparison
of the rate observed at the far-detector with such a no-oscillation rate determines
the lobe centered around such values of  $\Delta m^2_{14}$, and a second
less  pronounced (due to energy smearing) lobe around $\Delta m^2_{14} \sim 2\times 10^{-2}$\,eV$^2$,
corresponding to a phase a factor $2\pi$ larger in Eq.~(\ref{eq:phase}). 
At higher values of $\Delta m^2_{14}$, if one ideally had an anchor measurement 
at zero distance from the reactors, the exclusion contours would become vertical lines, 
corresponding to averaged oscillations at the far site. In the real situation, Bugey-4
is not located at zero distance and, as $\Delta m^2_{14}$  increases, the 4-flavor effects start 
to affect its total rate, which is no more a stable no-oscillation ``anchor''.  As a result, there is a gradual loss of sensitivity and no 
interesting limit can be put for $\Delta m^2_{14} \gtrsim 10^{-1}$\,eV$^2$.

The second panel of Fig.~2 represents the regions excluded by RENO. In this case,
the total rate measured by each of the two detectors is the sum of the partial contributions
arising from six reactors. In particular, from the geometry of the setup, one infers that each detector 
gets contributions from neutrinos that have traveled (and oscillated) along
three different baselines (see the figure at page 5 of~\cite{RENO_neutel}). Notwithstanding, 
the following qualitative observations can be done, making use of the concept
of  reactor-flux-weighted baselines, which for the near and far detectors are given respectively by $\sim0.4$~km and $\sim1.4$~km.
In the lower lobe centered around $\Delta m^2_{14} \sim 3 \times 10^{-2}$\,eV$^2$ the 4-flavor effects at 
the near detector are negligible, while the far detector has maximally sensitivity to the modulation
induced by the phase factor in Eq.~(\ref{eq:phase}). This behavior is very similar to that of Double Chooz
but here the lobe is shifted towards lower values of $\Delta m^2_{14}$ since the far-site baseline is longer
($\sim1.4$~km vs $\sim1.0$~km). The exclusion power is stronger compared to Double Chooz
because of the better precision of RENO. Finally, compared to Double Chooz, the lobe is broader,
as a result of the fact that the RENO far detector receives a weighted sum of neutrino fluxes
originating from reactors located at three different baselines, as noticed above. 
A second lobe is present around $\Delta m^2_{14} \sim 10^{-2}$\,eV$^2$.
In this region the total rate at the near detector is maximally sensitive to the modulation
of the oscillating factor in Eq.~(\ref{eq:phase}). On the other hand, the oscillations
at the far-detector are basically averaged with a total rate suppressed
by a $\Delta m^2_{14}$-independent factor, which provides a sort of anchor point
for the near detector. In the region of $\Delta m^2_{14}$ located above the two lobes the oscillations get averaged 
at both near and far detectors and their effects are invisible in the near-far comparison.
As a result, in this region there is no sensitivity to 4-flavor effects.

The third panel represents the regions excluded by Daya Bay. In this case a
quantitative interpretation is more complicated because in such an experiment 
there are six reactors and three detector halls. Two of them (EH1, EH2) mostly receive
neutrinos from the reactors positioned nearby, while the third one (EH3) receives neutrinos
from reactors located far apart (see Fig.~1 in~\cite{An:2012bu}). Qualitatively, we can observe that 
the two near detectors make possible a double near/far comparison, 
which is more informative than the single near/far comparison
made by RENO. By excluding from the analysis one of the two near halls
(EH1 or EH2) at a time, we have verified that each near/far ratio (EH1/EH3, EH2/EH3)
contributes to constrain the region $\Delta m^2_{14} \lesssim 4 \times 10^{-2}$\,eV$^2$.
In addition, by removing from the analysis the far detectors (located in EH3) we 
have checked that the ``near/near" ratio EH1/EH2 gives additional and complementary
constraints in the same region. Therefore, the three independent ratios confer to
Daya Bay a much more strong constraining power compared with RENO.

Finally, the fourth panel in Fig.~2 shows the combination
of the three experiments. Apart from the upper region around $\Delta m^2_{14} \gtrsim 4\times 10^{-2}$\,eV$^2$
which is excluded by the Double Chooz/Bugey-4 comparison, the rest of the plot
is dominated by Daya Bay.

\subsection{The general case of free $\theta_{13}$} 

Let us now come to the most general case in which $\theta_{13}$ is treated as a free parameter.
From the results displayed in Fig.~3, the following qualitative observations can be done.
I) As expected, for all  the three experiments the excluded regions are less restrictive
if compared to the case of fixed $\theta_{13}$ since
there is one additional parameter in the fit.  II) In certain specific ranges 
of $\Delta m^2_{14}$, which depend on the particular experiment,
there is a complete loss of sensitivity. This behavior is imputable to degeneracies
among $\theta_{13}$ and $\theta_{14}$, whose effects can cancel out in the near/far
ratios. In the case of Double Chooz such degeneracy problems are absent
because $\theta_{13}$ cannot mimic $\theta_{14}$ at the near site, where the 3-flavor 
effects are always negligible. At the contrary in RENO such effects are extremely important. 
In this case the entire lobe centered around $\Delta m^2_{14} \sim  10^{-2}$\,eV$^2$ disappears.
In Daya Bay the degeneracies are mitigated because, as discussed above, three distinct baselines (and rate ratios)
are probed. III) The impact of the degeneracies is almost neutralized in the global combination
due to a synergy of the multiple baselines comparison. When cancellation 
effects occur in one experiment, this does not happen for the the other two ones. 
As a non-trivial result, the combination of the three experiments is able to provide 
limits in all the region explored.

The natural question arises on the behavior of the estimate of $\theta_{13}$ in 
the presence of 4-flavor effects. Figure~4 answers this question by showing the 2-dimensional
projection in the space spanned by [$\sin^2\theta_{13}, \Delta m^2_{14}$], 
as determined after marginalization over the undisplayed variable $\theta_{14}$. 
Figure~4 shows that for $\Delta m^2_{14}\gtrsim 6\times 10^{-3}\,{\mathrm {eV}}^2$ 
the estimate of $\theta_{13}$ is quite robust and independent of the value of the
new mass-squared splitting. Its best fit value is always very close to that one obtained in the 3-flavor
case and also the statistical significance for its non-zero value is comparable. 
In the interval $\Delta m^2_{14}\in [1 - 4]\times 10^{-3}\,{\mathrm {eV}}^2$, which is centered around 
the atmospheric mass-squared splitting  ($\Delta m^2_{13} \simeq  2.4\times 10^{-3}\,{\mathrm {eV}}^2$),
the parameter $\theta_{13}$ has no lower bound. This behavior can be easily understood
as in this case the two terms in Eq.~(\ref{eq:pee_approx}), respectively driven by $\theta_{13}$ and $\theta_{14}$,
becomes almost identical and a complete degeneracy among the two parameters emerges.
In other words, for such values of $\Delta m^2_{14}$ all the reactor data could be interpreted in 
terms of pure $\theta_{14}$-driven oscillations without resorting to a non-zero $\theta_{13}$.
%%%%%%%%%%%%%%%%%%%%%%%%%%%%%%%%%%%%%%%%%%%%%%%%%
In practice, this interpretation is not possible due to the independent lower bound set on $\theta_{13}$ by
T2K~\cite{Abe:2013xua,T2K_EPS}, where 4-flavor effects%
%%%%%%%%%%%%%%%%%%%%%%%%%%%%%%%%%%%%%%%%%%%%%%%%%
\footnote{A discussion  (limited to the higher values of $\Delta m^2_{14} \sim 1\,$eV$^2$) of the 4-flavor effects in T2K 
can be found in~\cite{Bhattacharya:2011ee}.}  
%%%%%%%%%%%%%%%%%%%%%%%%%%%%%%%%%%%%%%%%%%%%%%%%%
are completely absent (for $U_{\mu4} = 0$) or have at most a subleading role (for a small non-zero value of $U_{\mu4}$).
Finally, for $\Delta m^2_{14} \lesssim 5\times 10^{-4}$\,eV$^2$ one recovers again the standard interval for $\theta_{13}$
since the phase in Eq.~(\ref{eq:phase}) becomes very small with any 4-flavor effect.

%%%%%%%%%%%%%%%%%%%%%%%%%%%%%%%%%%%%%%%%%%%
\begin{figure*}[t!]
\vspace*{-7.5cm}
\hspace*{1.00cm}
\includegraphics[width=20.0 cm]{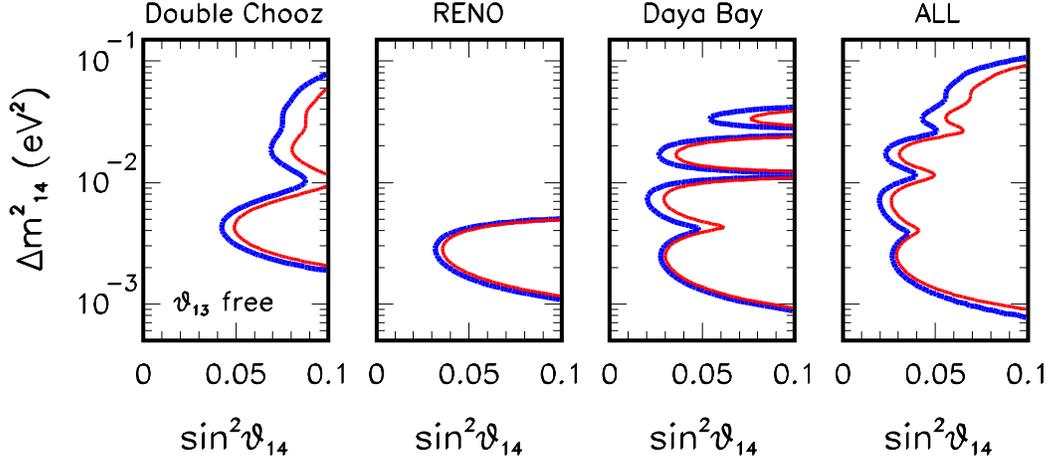}
\vspace*{-6.8cm}
\caption{Results obtained from the 4-flavor analysis for a free $\theta_{13}$, which is marginalized away. The contours refer to 2 d.o.f. 90\% C.L. (blue thick line) and 99\% (red thin line).
\label{fig3}}
\end{figure*}  
%%%%%%%%%%%%%%%%%%%%%%%%%%%%%%%%%%%%%%%%%%%

%%%%%%%%%%%%%%%%%%%%%%%%%%%%%%%%%%%%%%%%%%%%%
\begin{figure*}[b!]
\vspace*{-1.2cm}
\hspace*{2.50cm}
\includegraphics[width=12.0 cm]{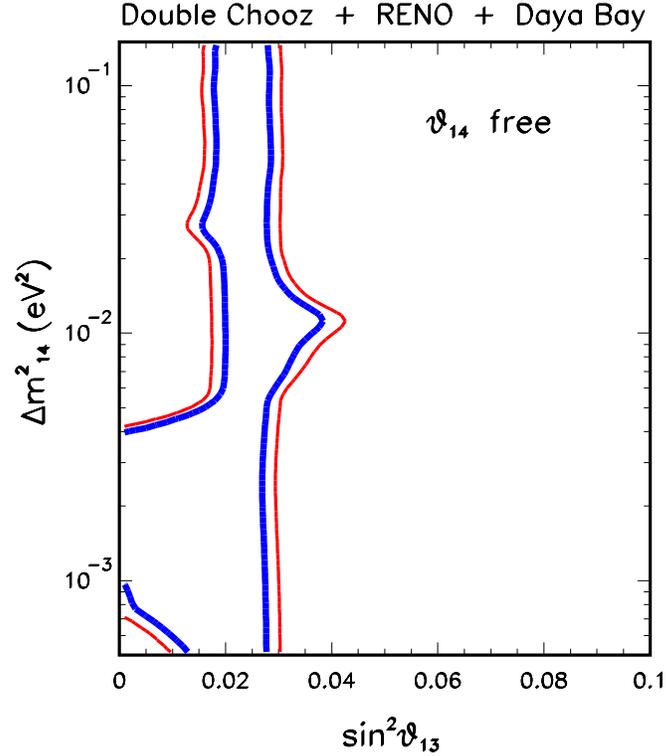}
\vspace*{-0.5cm}
\caption{Regions allowed by the combination of the three reactor experiments.
The undisplayed parameter   $\theta_{14}$ is marginalized away.
The contours refer to 90\% (blue thick line) and 99\% (red thin line) for 2 d.o.f.
\label{fig4}}
\end{figure*}  
%%%%%%%%%%%%%%%%%%%%%%%%%%%%%%%%%%%%%%%%%%%%%

\section{Conclusions and discussion}

We have performed a detailed study of the limits attainable from the three
$\theta_{13}$-sensitive reactor experiments Double Chooz, RENO and Daya Bay
on very light sterile neutrinos.
We have shown that the multi-baseline total-rate comparison provided by such
experiments makes it possible to obtain interesting constraints on the mixing
matrix element $|U_{e4}|^2$  for mass-squared splittings in the range
$[10^{-3} - 10^{-1}]$~eV$^2$. 
In addition, our analysis shows that the reactor estimate of $\theta_{13}$ is robust and substantially
independent of the 4-flavor-induced perturbations provided that  new
mass-squared splitting  is not too low ($\gtrsim 6 \times10^{-3}\,\mathrm{eV}^2$). 
Due to a lack of information we have restricted our analysis to the total rates.
We expect a substantial improvement of our limits with the inclusion
of the spectral shape information, whose details, we hope, may become publicly available
in the near future. Equally important will be the clarification of the ``reactor anomaly'',
as it would allow us to use the information on the absolute neutrino fluxes, which was 
``integrated out'' in our analysis. In this respect, the publication of the absolute neutrino flux
measured by Daya-Bay and RENO will be very important.

While we have limited our study to reactor neutrinos, which are sensitive to the 
electron neutrino mixing $U_{e4}$, complementary information on VLS$\nu$'s may be
obtained from other kinds of experiments. In the explored range of mass-squared
splittings one may expect subleading effects in long-baseline accelerator setups
and in atmospheric neutrinos, which are sensitive to $L/E_\nu$ values similar to those
explored in $\theta_{13}$-sensitive reactor experiments. In particular,
one may constrain the matrix element $|U_{\mu4}|$ by looking at the $\nu_\mu \to \nu_\mu$ 
disappearance channel (MINOS~\cite{Adamson:2011ig}, T2K~\cite{Abe:2013fuq}, and atmospheric data) 
and at the MINOS neutral current measurements~\cite{Adamson:2011ku} and probe
the product $|U_{e4}||U_{\mu4}|$ by looking at the appearance $\nu_\mu \to \nu_e$ channel
(T2K~\cite{Abe:2013xua,T2K_EPS}, ICARUS~\cite{Antonello:2013gut}, OPERA~\cite{Agafonova:2013xsk} and atmospheric data).

Finally, we would like to underline that the region of mass-mixing parameters
not excluded by our analysis is of great interest for cosmology. In fact, for mass-squared splittings in the range $[10^{-3} - 10^{-1}]$~eV$^2$,
and sufficiently small admixtures  $|U_{e4}|^2 \lesssim {\mathrm {few}} \times 10^{-2}$,
a fourth sterile neutrino is only partially thermalized in the early universe and thus provides
a fractional contribution $\Delta N_{\mathrm {eff}} \in [0, 1]$ to the number of effective 
relativistic degrees of freedom (see~\cite{Hannestad:2012ky,Mirizzi:2013kva}), as indicated by the latest
cosmological measurements~\cite{Ade:2013zuv}, which give $\Delta N_{\mathrm {eff}} = 0.62^{ +0.50}_{-0.48}$
(95\% C.L.). Moreover, the absolute neutrino mass content implied by a 3+1 scheme involving a VLS$\nu$  is
very small and fully compatible with the existing (sub-eV) upper limits (see for example~\cite{Giusarma:2013pmn}).
Therefore, a very light sterile neutrino is a credible candidate to explain the ``dark radiation'' excess and 
 the preference for a hot dark matter component with particle mass in the sub-eV range recently emerged
 in cosmological data analyses~\cite{Wyman:2013lza,Hamann:2013iba,Burenin:2013wg}.
These circumstances reinforce the motivation for a vigorous program of
investigation aimed at improving the sensitivity to VLS$\nu$ mass-mixing parameters and
at assessing their impact in all the astroparticle ``laboratories''.

\section*{Acknowledgments}

We thank Soo-Bong Kim for providing useful information on the RENO setup.
We acknowledge support from the European Community through a Marie Curie IntraEuropean 
Fellowship, grant agreement no. PIEF-GA-2011-299582, ``On the Trails of New Neutrino Properties".
We also acknowledge partial support from the European Union FP7 ITN Invisibles (Marie Curie Actions, 
PITN-GA-2011-289442).

\end{document}